\documentclass[10pt,doublecolumn]{IEEEtran}
\usepackage{amsmath,amsthm,amsfonts,amssymb,bm,mathrsfs}
\usepackage{graphicx,subfigure}
\usepackage[bookmarks=false,colorlinks,citecolor=black, filecolor=black, linkcolor=black,urlcolor=black]{hyperref}
\usepackage{url}
\usepackage[cmintegrals]{newtxmath}
\usepackage{amssymb}

\ifCLASSINFOpdf
\else
\fi
\begin{document}
	
%
\title{On the Capacity of Fractal Wireless Networks With Direct Social Interactions}

\author{Ying Chen, \IEEEauthorblockN{Rongpeng Li, Zhifeng Zhao, and Honggang Zhang}\\
	\IEEEauthorblockA{York-Zhejiang Lab for Cognitive Radio and Green Communications}\\
	\IEEEauthorblockA{College of Information Science and Electronic Engineering\\
		Zhejiang University, Zheda Road 38, Hangzhou 310027, China\\}
	\thanks{Y. Chen, Z. Zhao, R. Li, and H. Zhang are with College of Information Science and Electronic Engineering, Zhejiang University. Email: \{21631088chen\_ying, lirongpeng, zhaozf, honggangzhang\}@zju.edu.cn}
	\thanks{This paper is financially supported by the Program for Zhejiang Leading Team of Science and Technology Innovation (No. 2013TD20), the Zhejiang Provincial Technology Plan of China (No. 2015C01075), and the National Postdoctoral Program for Innovative Talents of China (No. BX201600133).}}


%


\maketitle

\begin{abstract}
 The capacity of a fractal wireless network with direct social interactions is studied in this paper. Specifically, we mathematically formulate the self-similarity of a fractal wireless network by a power-law degree distribution $ P(k) $, and we capture the connection feature between two nodes with degree $ k_{1} $ and $ k_{2} $ by a joint probability distribution $ P(k_{1},k_{2}) $. It is proved that if the source node communicates with one of its direct contacts randomly, the maximum capacity is consistent with the classical result $ \Theta\left(\frac{1}{\sqrt{n\log n}}\right) $ achieved by Kumar \cite{Gupta2000The}. On the other hand, if the two nodes with distance $ d $ communicate according to the probability $ d^{-\beta} $, the maximum capacity can reach up to $ \Theta\left(\frac{1}{\log n}\right) $, which exhibits remarkable improvement compared with the well-known result in \cite{Gupta2000The}.
\end{abstract}



\section{Introduction}
The research on the capacity of a wireless network has aroused intense interests in recent years. However, little attention has been paid to fractal wireless networks. As a vital property of networks, fractal behavior has been discovered in many wireless networking scenarios \cite{Yuan2016Not}. For example, the coverage boundary of the wireless cellular networks shows a fractal shape, and the fractal feature can inspire the design of the handoff scheme in mobile terminals \cite{Ge2016Wireless}. Moreover, many significant networks in the real world exhibit the fractal characteristics naturally, such as the protein homology network, the undirected WWW and so on \cite{Gallos2008Scaling}. 

In the theory of fractal wireless networks, self-similarity belongs to one of the most important characteristics. Self-similarity of fractal wireless networks requires the node degree distribution $ P(k) $ to remain unchanged when the network grows. In order to meet this requirement, power-law degree distribution $ P(k)\sim k^{-\gamma} $ must be followed in a fractal wireless network. Actually, the power-law distribution lays the foundation for the capacity analysis of fractal wireless networks.

In $ 2000 $, Gupta and Kumar proved that the throughput in wireless networks can only reach $ \Theta\left(\frac{1}{\sqrt{n\log n}}\right) $ when the network size is $n$ \cite{Gupta2000The}, where the symbol $ \Theta $ refers to the order of magnitude and $ T(n)=\Theta (h(n)) $ means that the two functions $T(n)$ and $h(n)$ have the same rate of growth. Afterwards, H. R. Sadjadpour et al. discovered that the maximum capacity can be improved in scale-free networks \cite{Azimdoost2013Capacity}\cite{Kiskani2013Social}\cite{Karimzadeh2011Effect}. However, to our best knowledge, there have been no theoretical researches about the capacity of fractal wireless networks with direct social interactions, which is exactly what we study in this paper.

In the remainder of this article, we introduce the foundations of a fractal wireless network and discuss  the  network model in Section II. Then the maximum throughput is derived in Section III and the results are discussed in Section IV. Finally, we draw a conclusion in Section V.

\section{Background and Models}
\subsection{Foundations of A Fractal Network}
Fig.1 illustrates the topology in depth of a general fractal wireless network, in which we assume that a node $ v_{i} $ is going to transmit a data packet to another node. In other words, the node $ v_{i} $ is the source node. The nodes $ v_{j1} $ and $ v_{j2} $ are directly connected with the node $ v_{i} $ and are regarded as the direct social contacts of $ v_{i} $. Then $ v_{k} $ is chosen among the direct social contacts to communicate with $ v_{i} $ and it is known as the destination node. Usually, a node has many direct social contacts, and the degree of one node refers to the number of its direct social contacts. 
\begin{figure}[htbp]
	\centering
	\includegraphics[scale=0.5]{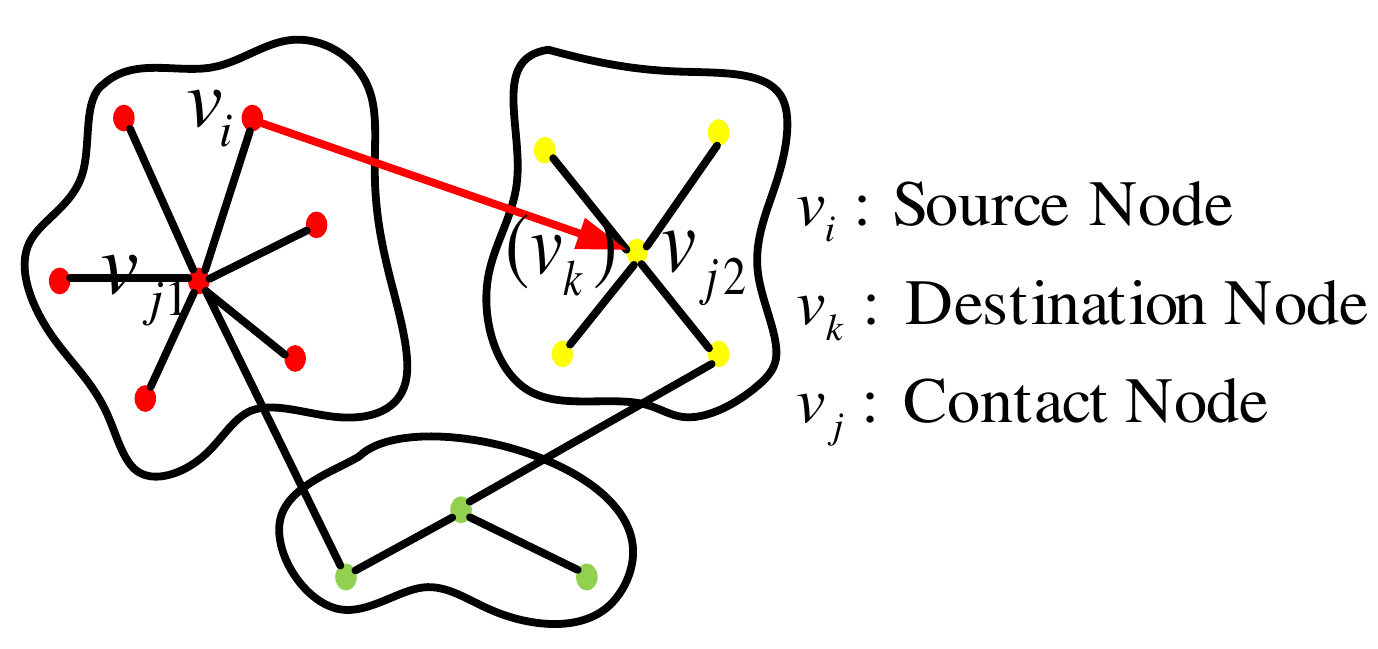}
	\caption{The topology illustration of a fractal wireless network.}
\end{figure}

In order to characterize fractal wireless networks, it is essential to introduce the concept of renormalization through the box-covering algorithm \cite{Song2006Origins}. Renormalization is a technique to examine the internal relationship by using a box to cover several nodes and virtually replacing the whole box by a new node. Besides, if there exists a link between two nodes in two boxes respectively, then the two corresponding nodes evolved from the boxes will be connected. Mathematically, the network could be minimally  covered by $ N_{B}(l_{B}) $ boxes of the same length scale $ l_{B} $ under the renormalization, where $ l_{B} $ is the size of boxes measured by the maximum path length between any pair of nodes inside the box and $ N_{B}(l_{B}) $ is the minimum value among all possible situations. If the network is a fractal wireless network, we have the following relations, namely \cite{Song2006Origins}:
\[\left\{ {\begin{array}{*{20}{c}}
\begin{aligned}
&{{N_B}({l_B})\sim n \cdot {l_B}^{ - {d_B}}}\\
&{{k_B}({l_B})/{k_{hub}}\sim {l_B}^{ - {d_g}}}\\
&{{n_h}({l_B})/{k_B}({l_B})\sim {l_B}^{ - {d_e}}},
\end{aligned}
\end{array}} \right.\]

\noindent where $ n $ is the number of nodes in the network. A hub indicates the node of the largest degree inside each box, while $ k_{B}(l_{B}) $ and $ k_{hub} $ denote the degree of the box and the hub respectively. $ n_{h}(l_{B}) $ refers to the number of links between the hub of a box and nodes in other connected boxes. The three indexes $ d_{g} $, $ d_{B} $ and $ d_{e} $ denote the degree exponent, the fractal exponent, and the anti-correlation exponent, respectively.

In addition, for fractal wireless networks, the degree exponent $ d_{g} $ and the fractal exponent $ d_{B} $ are both finite. The anti-correlation exponent $ d_{e} $ reveals the repulsion effect between the hubs while large anti-correlation exponent tends to result in fractal wireless networks.

On the other hand, the joint probability distribution $P(k_1,k_2)$ gauges the possibility to have a connection between two nodes with degree $ k_{1} $ and $ k_{2} $. In fractal wireless networks, $P(k_1,k_2)$ can be expressed as \cite{Gallos2008Scaling}:
\[ P(k_1,k_2)\sim k_1^{-(\gamma-1)}\cdot k_2^{-\epsilon} (k_{1}>k_{2}),
		\]

\noindent where $ \gamma $ is the degree distribution exponent and $ \epsilon $ is the correlation exponent. It is proved that there exist certain relations among the aforementioned key parameters: $ \gamma  = 1 + \frac{{{d_B}}}{{{d_g}}} $ and $ \epsilon  = 2 + \frac{{{d_e}}}{{{d_g}}} $ \cite{Gallos2008Scaling}, which suggest that $ \gamma $ and $ \epsilon $ in a fractal wireless network are larger than $ 1 $ and $ 2 $ respectively. A reasonable explanation for the condition $ k_{1}>k_{2} $ is that nodes with larger degrees have a higher priority in contacts selection, so they choose nodes of smaller degrees and set up connections with them. 

\subsection{Network Model}
In order to clarify the capacity of a fractal wireless network with direct social interactions clearly and orderly, we assume that the fractal social network is deployed in a unit area square and the $ n $ nodes are uniformly distributed, as displayed in Fig. 2.
{\setlength\abovecaptionskip{0.cm}
	\setlength\belowcaptionskip{-10.cm}\begin{figure}
	\centering
	\includegraphics[scale=0.3]{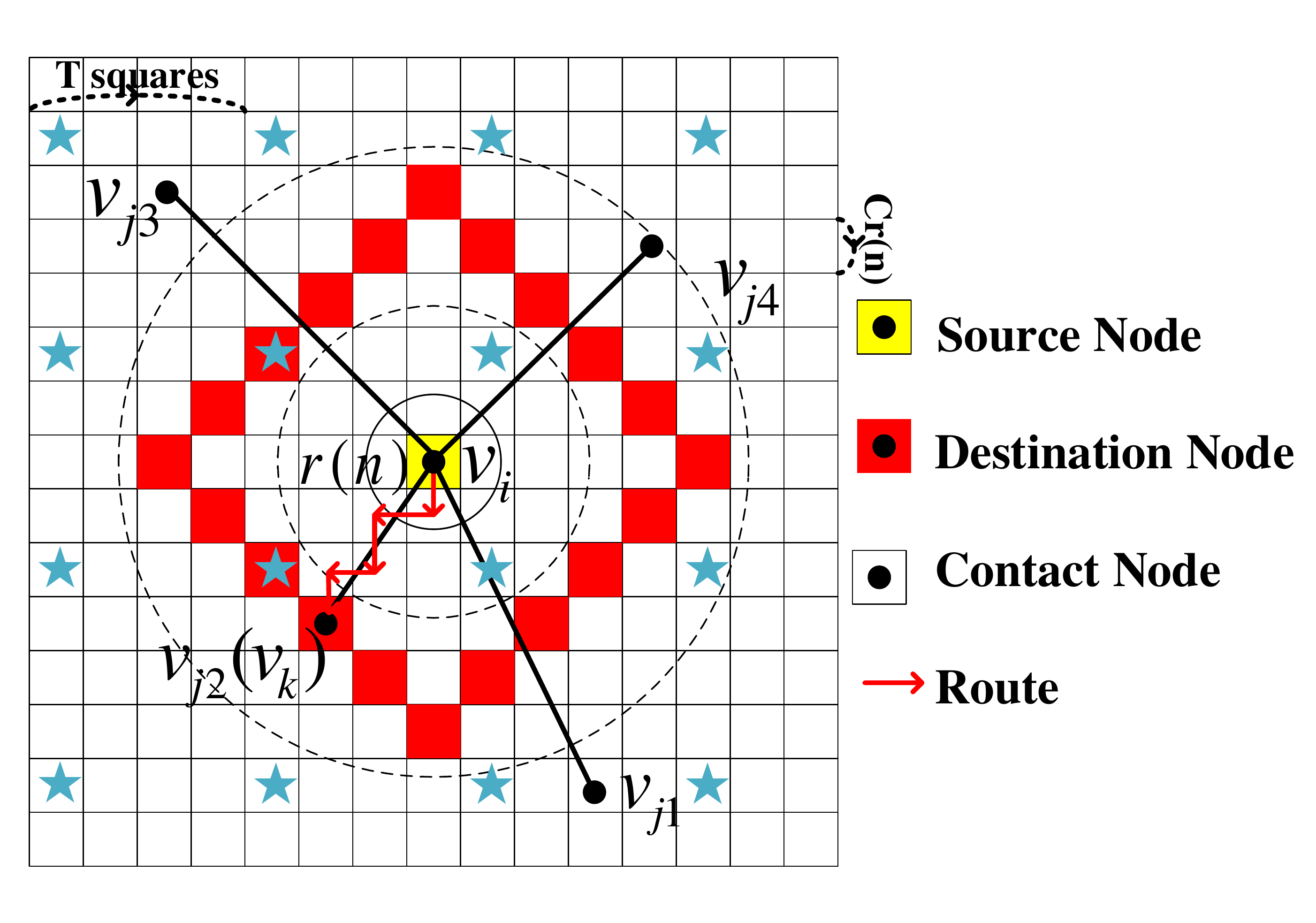}
	\caption{The fractal network in a standard unit area square model.}
\end{figure}

The Protocol Model in \cite{Gupta2000The} is adopted as the measurement of a successful transmission. A transmission is successful if and only if $ |X_{i}-X_{j}|\leqslant r(n) $ and $ |X_{k}-X_{i}|\geq (1+\Delta)|X_{i}-X_{j}| $. $ X_{i} $ and $ X_{j} $ refer to the transmitter node and the receiver node respectively. $ X_{k} $ denotes any other transmitter sharing the same channel with $ X_{i} $ and $ \Delta $ is the protection factor of interference zone. To eliminate the isolated nodes, the transmission range $ r(n) $ has to exceed some threshold. It is proved that $ r(n) $ must reach $ \Theta\left(\sqrt{\frac{\log n}{n}}\right) $ to guarantee the connectivity of the network \cite{Gupta2000The}. In Fig. 2, the  solid-line circle with the radius of $ r(n) $ displays the transmission range.

A TDMA (Time Division Multiple Access) scheme is designated as the MAI (multiple access interference) avoidance method in the time domain. The network is cut into a number of smaller squares with side length $ C_{1}r(n) $. The interference units refer to those cells containing at least two nodes closer than $ (2+\Delta)r(n) $ respectively[1], and cells, which simultaneously transmit, should not be the interference units with each other. Therefore, the squares signed with stars, whose distances are at least $ T $ squares from each other, are permitted to transmit at the same time, where $ T\geq (2+\Delta)/C_{1} $.

Since the selection of the destination node affects the capacity of a fractal wireless network as well, in this paper, the destination node is chosen in two different ways. In the first case, the destination node is selected according to the uniform distribution $ \frac{1}{q} $. That is to say, the source node communicates with one of its direct social contacts randomly, and all the contacts have the same opportunities to communicate with the source node. In another case, it is considered to be reasonable that the destination node is selected according to the power-law distribution $ d^{-\beta} $\cite{Latane1995Distance}, where $ d $ refers to the distance between the source node and the destination node, and $ \beta $ indicates that the source node communicates with different nodes with different preferences\cite{Karimzadeh2011Effect}. In other words, the direct social contacts closer to the source node have more opportunities to communicate with it. On the other hand, it is noteworthy that $ \beta $ is closely related with the configuration of the fractal wireless networks. Accordingly, the parameter $ \beta $ can be noted as $ \beta=\beta(\gamma,\epsilon) $, which indicates that $ \beta $ is a function of the degree distribution exponent $ \gamma $ and the correlation exponent $ \epsilon $. 

The routing scheme in the space domain is pretty simple. When the source node is about to send a data packet, it chooses one node from its neighbor squares closest to the destination node to relay the packet. The packet eventually reaches the destination node after multiple hops. The red line with arrows denotes one possible routing path in Fig. 2. All the squares in red have the same hops $ x $ from the source node and the total number of them is $ 4x $. The radius of the two dotted-line circles are used as the distances between the nodes in red squares and the source node instead of their real distances. 

$ \mathbf{Definition\ 1.} $ The elementary symmetric polynomial $ \sigma_{p,N}(Q)$, $1\leq p\leq N $ \cite{JIPAM2003New} of variables $ Q=(q_{1},q_{2},\ldots,q_{N}) $ is noted as
\[ 
\begin{array}{l}
\sigma_{p,N}(Q)=\sigma_{p,N}(q_{1},q_{2},\ldots,q_{N})\\
\quad \quad \qquad=\sum\nolimits_{ 1\leq i_{1}\leq i_{2}\leq\ldots i_{p}\leq N}q_{i_{1}}q_{i_{2}}\cdots q_{i_{p}}.
\end{array}
 \]

$ \mathbf{Definition\ 2.} $ The elementary symmetric polynomial $ \sigma^{\overline{k}}_{p,N-1}(Q) $, $ 1\leq p\leq N-1 $ of variables $ Q=(q_{1},q_{2},\ldots,q_{N}) $ except $ q_{k} $ is noted as
\[
\sigma^{\overline{k}}_{p,N-1}(Q)=\sigma_{p,N-1}(q_{1},q_{2},\ldots,q_{k-1},\ldots,q_{k+1},\ldots,q_{N}).
\]

 $ \mathbf{Lemma\ 1.} $ Let the set $ Q = \{ {q_1},{q_2}, \cdot  \cdot  \cdot ,{q_N}\}  $ contains $ N\geq 2 $ non-negative real numbers. If $ q $ is finite, then we have \cite{Karimzadeh2011Effect}
\[ 
	\frac{\sigma_{1,N}(Q)\sigma_{q,N}(Q)}{(q+1)\sigma_{q+1,N}(Q)}=\Theta(\frac{N}{N-q}).
\]

\section{The Upper Bound of the Capacity}

In this section, we will follow the aforementioned properties of fractal wireless networks and try to clarify the specific derivation procedure of the maximum capacity with direct social interactions. In addition, we also study the impact of a particular destination selection rule on the maximum achievable throughput by taking account of two different cases, which include uniformly and power-law distributed destinations.  

\subsection{The Case of Uniformly Distributed Destinations}
In the first case, we take the uniform distribution of the destination node into consideration. In other words, we assume that the source node selects one of its direct social contacts as the destination node randomly, so each contact node has the same possibility to communicate with the source node. In this situation, we give the result of the maximum capacity in Theorem 1 and proof it subsequently.

$ \mathbf{Theorem\ 1.} $ For a fractal wireless network with $ n $ nodes satisfying the conditions below: 1) the direct social contacts are selected according to the joint probability distribution $ P({k_1},{k_2}) = \frac{{k_1^{ - (\gamma  - 1)}k_2^{ - \epsilon }}}{{{M_{\gamma ,\epsilon }}}} $; 2) the degree of each node follows the power-law degree distribution $ P(k) = \frac{{{k^{ - \gamma }}}}{{\sum_{k = 1}^n {{k^{ - \gamma }}} }} $; 3) the destination node is chosen by the uniform distribution $ P({v_t} = {v_k}|{v_k} \in {\bf{C}}) = \frac{1}{q} $.  Then the maximum capacity $ \lambda_{max} $ of the fractal wireless network with direct social interactions is
\[{\lambda _{\max }} = \Theta\left(\frac{1}{{\sqrt {n \cdot \log n} }}\right).
\]

In order to get the result in Theorem 1, we first present the relationship between the capacity and the average number of hops $ E[X] $, so we can solve the problem by finding out $ E[X] $. Secondly, we give the expression of $ E[X] $. Thirdly, We divide $ E[X] $ into two separate cases $ E_{1} $ and $ E_{2} $ with a boundary $ q_{0} $, which indicates a relatively large degree. $ E_{1} $ is the average number of hops when $ q \le {q_0} $, where the degree $ q $ of the source node is a finite integer, meanwhile $ E_{2} $ is the average number of hops when $ q > {q_0} $, where $ q $ is considered to be infinite. After $ E_{1} $ and $ E_{2} $ are obtained respectively, the capacity derivation is achieved by backtracking. Next, we will follow the proof sketch above, and then have the following lemmas.  

$ \mathbf{Lemma\ 2.} $ Assume that $ \lambda $ is the data rate for every node, $ X $ is the number of hops from the the source node to the destination node. $ E[X] $ denotes the expectation of $ X $ for any transmission between a pair of source and destination nodes. $ W $ is the bandwidth assigned to the whole network. Then we have 
\begin{equation}\label{1}
\lambda\leqslant\lambda_{max}=\Theta\left(\frac{1}{\log n\cdot E[X]}\right).
\end{equation}

Proof: The total traffic which the network has to deal with is $ n\lambda E[X] $ while the network can handle $ \frac{W}{T^{2}C_{1}^{2}r(n)^{2}} $ simultaneously. So the inequality $ n\lambda E[X] \le \frac{W}{{{T^2}{C_{1}^2}{r^2}(n)}} $ must be satisfied. Recalling that the order of $ r(n) $ is $ \Theta\left(\sqrt{\frac{\log n}{n}}\right) $ in Section II, then we can get Eq. \eqref{1}. $\hfill\blacksquare$ 

$ \mathbf{Lemma\ 3.} $ Let the source node $ v_{i} $ locate in the central square whose degree is $ q $, where $ q=1,2,\ldots,n $. $ v_{k} $ denotes the destination node whose degree is $ q_{k} $. The variables $ {q_1},{q_2}, \cdot  \cdot  \cdot ,{q_N} $ in $ Q = (q_1^{ - \epsilon },q_2^{ - \epsilon }, \cdot  \cdot  \cdot ,q_N^{ - \epsilon }) $ denote the degrees of $ N $ potential direct social  contacts whose degrees are smaller than $ q $. Then the average number of hops can be given as 
\begin{equation}\label{2}
E[X] = \sum\limits_{q = 1}^n {\frac{{{q^{ - \gamma }}}}{{\sum_{b = 1}^n {{b^{ - \gamma }}} }}}  \cdot \sum\limits_{x=1}^{\frac{1}{{r(n)}}} {x\sum\limits_{l = 1}^{4x} {\sum\limits_{{v_k} \in {s_l}} {\frac{{q_k^{ - \epsilon }\sigma _{q - 1,N - 1}^{\overline k }(Q)}}{{{q\cdot \sigma _{q,N}}(Q)}}} } }.
\end{equation}

Proof: Let $ P(k=q) $ denote the probability when the degree of the source node is $ q $, while $ E[X|{\rm{source  \ }}{v_i},k = q] $ is the average number of hops under the condition that the source node $ v_{i} $ has $ q $ direct social contacts, then $ E[X] $ can be written as
\begin{equation}\label{3}
E[X] = \sum\limits_{q=1}^{n} P(k=q) E[X|{\rm{source\ }}{v_i},k = q].
\end{equation}

Let $ P(X=x) $ denote the probability of $ x $ hops ranging from 1 to $ 1/r(n) $. The event $ X=x $ is true if and only if $ v_{k} $  locates in the red squares $ s_{l}(l=1,2,\ldots 4x) $ and it is selected as the destination node $ v_{t} $. Therefore, $ E[X|{\rm{source \ }}{v_i},k = q] $ can be expanded as
\begin{equation}\label{4}
\begin{array}{l}
E[X|{\rm{source\ }}{v_i},k = q] = \sum\limits_{x=1}^{\frac{1}{r(n)}} {x \cdot } P(X = x)\\
\quad \qquad\qquad\qquad\qquad= \sum\limits_{x=1}^{\frac{1}{r(n)}} {x \cdot } \sum\limits_{l = 1}^{4x} {\sum\limits_{{v_k} \in {s_l}} {P({v_t} = {v_k})} }.
\end{array}
\end{equation}

Let $ \mathbf{C} $ be the set of all direct social contacts. $ v_{t}=v_{k} $ implies that $ v_{k} $ is chosen as the destination node $ v_{t} $ after it is selected as a direct social contact. In other words,
\begin{equation}\label{5}
P(v_{t}=v_{k})=P(v_k\in\mathbf{C})\cdot P(v_t=v_k|v_k\in\mathbf{C}).
\end{equation}

The set $ \{v_{i_{1}},v_{i_{2}},\ldots,v_{i_{q}}\} $ contains $ q $ direct social contacts of the source node. Taking all possible combinations of nodes into consideration, the probability that the source node has $ q $ direct social contacts is
\[ \begin{array}{ll}
P(|\mathbf{C}|=q)=\sum\nolimits_{1\leq i_{1}\leq i_{2}\leq\ldots i_{q}\leq N}P(\mathbf{C}=\{v_{i_{1}},v_{i_{2}},\ldots,v_{i_{q}}\})\\
\qquad\qquad\quad =\sum\nolimits_{1\leq i_{1}\leq i_{2}\leq\ldots i_{q}\leq N} \frac{(q^{-(\gamma-1)})^{q}q^{-\epsilon}_{i_{1}}q^{-\epsilon}_{i_{2}}\cdots q^{-\epsilon}_{i_{q}}}{(M_{\gamma,\epsilon})^{q}},	
\end{array} \]

\noindent where $ N $ is the number of nodes whose degrees are less than $ q $ and the source node selects direct social contacts only among those  nodes. $ N $ grows as fast as $ n $ because
\[ 
N=N_{<q}=n\cdot \frac{\sum^{q-1}_{b=1}b^{-\gamma}}{\sum^{n}_{b=1}b^{-\gamma}}=\Theta(n).
\]

The probability that the set $ \mathbf{C} $ consists of $ q $ particular nodes is
\[P({\mathbf{C}} = {\rm{\{ }}{v_{{i_1}}},{v_{{i_2}}}, \cdot  \cdot  \cdot {v_{{i_q}}}\} {\rm{)}} = \frac{{q_{{i_1}}^{ - \epsilon }q_{{i_2}}^{ - \epsilon } \cdot  \cdot  \cdot q_{{i_q}}^{ - \epsilon }}}{{\sum_{1 \le {i_1} \le  \cdot  \cdot  \cdot  \le {i_q} \le N} {{\rm{    }}q_{{i_1}}^{ - \epsilon }q_{{i_2}}^{ - \epsilon } \cdot  \cdot  \cdot q_{{i_q}}^{ - \epsilon }} }}.
\]

Consequently, we obtain the probability that $ v_{k} $ is selected as a direct social contact and simplify it with the elementary symmetric polynomials in Definition 1 and 2, namely:
\begin{equation}\label{6}
\begin{array}{l}
P({v_k} \in {\bf{C}}{\rm{)}} = \frac{{q_k^{ - \epsilon }\sum_{1 \le {i_1} \le  \cdot  \cdot  \cdot  \le {i_{q - 1}} \le N} {{\rm{    }}q_{{i_1}}^{ - \epsilon }q_{{i_2}}^{ - \epsilon } \cdot  \cdot  \cdot q_{{i_{q - 1}}}^{ - \epsilon }} }}{{\sum_{1 \le {i_1} \le  \cdot  \cdot  \cdot  \le {i_q} \le N} {{\rm{    }}q_{{i_1}}^{ - \epsilon }q_{{i_2}}^{ - \epsilon } \cdot  \cdot  \cdot q_{{i_q}}^{ - \epsilon }} }}\\
\qquad\qquad\ = \frac{{q_k^{ - \epsilon }\sigma _{q - 1,N - 1}^{\overline k }(Q)}}{{{\sigma _{q,N}}(Q)}}.
\end{array}
\end{equation}

Then we have the complete expression of Eq. \eqref{2} after expanding Eq. \eqref{3} with Eq. \eqref{4} - Eq. \eqref{6}. $\hfill\blacksquare$ 

$ \mathbf{Lemma\ 4.} $ When the degree of the source node is not greater than $ q_{0} $, i.e. $ q \le {q_0} $, the average number of hops $ E_{1} $ is
\begin{equation}\label{7}
E_{1} =\Theta \left(r{{(n)}^{ - 1}}\right). 
\end{equation}

Proof: According to the meaning of $ E_{1} $, it should be given as
\begin{equation}\label{8}
{E_1} = \sum\limits_{q = 1}^{{q_0}} {\frac{{{q^{ - \gamma }}}}{{\sum_{b = 1}^n {{b^{ - \gamma }}} }}}  \cdot \sum\limits_{x=1}^{\frac{1}{{r(n)}}} {x\sum\limits_{l = 1}^{4x} {\sum\limits_{{v_k} \in {s_l}} {\frac{{q_k^{ - \epsilon }\sigma _{q - 1,N - 1}^{\overline k }(Q)}}{q\cdot{{\sigma _{q,N}}(Q)}}} } } .
\end{equation}

All situations of selecting $ q-1 $ nodes from $ \mathbf{C} $ can be parted into two categories according to the condition whether $ {v_k} $ is chosen. If it is chosen, we have to select the other $ q-2 $ nodes from $ \mathbf{C} $ except $ {v_k} $. Otherwise we select the other $ q-1 $ nodes in $ \mathbf{C} $ except $ {v_k} $, that is to say,
\[\begin{array}{l}
\quad \sigma _{q - 1,N - 1}^{\overline k }(Q) = {\sigma _{q - 1,N}}(Q) - q_k^{ - \epsilon }\sigma _{q - 2,N - 1}^{\overline k }(Q){\rm{ }}\\
{\rm{                  }} = {\sigma _{q - 1,N}}(Q) - q_k^{ - \epsilon }({\sigma _{q - 2,N}}(Q) - q_k^{ - \epsilon }\sigma _{q - 3,N - 1}^{\overline k }(Q)).
\end{array}\]

Since every term in the equation above is positive, then we have
\begin{equation}\label{9}
{\sigma _{q - 1,N}}(Q) - q_k^{ - \epsilon } \cdot {\sigma _{q - 2,N}}(Q) < \sigma _{q - 1,N - 1}^{\overline k }(Q) < {\sigma _{q - 1,N}}(Q).
\end{equation}

A transformation of the Lemma 1 suggests that
\begin{equation}\label{10}
\frac{{{\sigma _{1,N}}(Q){\sigma _{q - 1,N}}(Q)}}{{q \cdot {\sigma _{q,N}}(Q)}} = \Theta (\frac{N}{{N - q + 1}}) = \Theta (1).
\end{equation}

Moreover, the probability that the degree of the source node is not greater than $ q_{0} $ is
\begin{equation}\label{11}
P(q \le {q_0}) = \sum\limits_{q = 1}^{{q_0}} {\frac{{{q^{ - \gamma }}}}{{\sum_{b = 1}^n {{b^{ - \gamma }}} }}}  = \Theta (1).
\end{equation}

Therefore, we get the upper bound of $ E_{1} $ according to Eq. \eqref{8} - Eq. \eqref{11}.
\begin{equation}\label{12}
\begin{array}{l}
E_{1} < \sum\limits_{x=1}^{\frac{1}{{r(n)}}} {x\sum\limits_{l = 1}^{4x} {\sum\limits_{{v_k} \in {s_l}} {\frac{{q_k^{ - \epsilon }{\sigma _{q - 1,N}}(Q)}}{q\cdot{{\sigma _{q,N}}(Q)}}} } } \\
\quad \equiv \frac{{q_k^{ - \epsilon }}}{q\cdot{{\sigma _{1,N}}(Q)}}\sum\limits_{x=1}^{\frac{1}{{r(n)}}} {x\sum\limits_{l = 1}^{4x} {\sum\limits_{{v_k} \in {s_l}} 1 } }. 
\end{array}
\end{equation}

\noindent where the symbol $ \equiv $ indicates the same order of magnitude on the two sides of an equation.

All the $ n $ nodes are distributed uniformly in the unit area, and the side length of each square is $ C_{1}r(n) $, so the summation term in Eq. \eqref{12} can be solved as
\begin{equation}\label{13}
\begin{array}{l}
\sum\limits_{x=1}^{\frac{1}{{r(n)}}} {x\sum\limits_{l = 1}^{4x} {\sum\limits_{{v_k} \in {s_l}}1 } }\\
\equiv \sum\limits_{x=1}^{\frac{1}{{r(n)}}} {x \cdot 4x \cdot {C_{1}^2}{r^2}(n)}  \cdot n \cdot 1{\rm{ }}\\
\equiv n \cdot r{(n)^2}\sum\limits_{x=1}^{\frac{1}{{r(n)}}} {{x^2} {\rm{ }}}\\
\equiv \Theta \left(n\cdot r{{(n)}^{ - 1}}\right).
\end{array}
\end{equation}

The $ q_k^{ - \epsilon } $ term in Eq. \eqref{12} can be replaced with its mean in the upper bound for convenience,
\begin{equation}\label{14}
E[q_k^{ - \epsilon }] = \sum\limits_{b = 1}^{q - 1} {P(k = b) \cdot {b^{ - \epsilon }}}  = \frac{{\sum_{b = 1}^{q - 1} {{b^{ - (\gamma  + \epsilon )}}} }}{{\sum_{b = 1}^n {{b^{ - \gamma }}} }} \equiv \Theta (1).
\end{equation}

On the other hand,
\begin{equation}\label{15}
{\sigma _{1,N}}(Q) = \sum\limits_{j = 1}^N {q_j^{ - \epsilon } \equiv N \cdot \int_1^{q - 1} {{u^{ - \epsilon }}\frac{{{u^{ - \gamma }}}}{{\sum_{b = 1}^n {{b^{ - \gamma }}} }}} } du \equiv \Theta (n),
\end{equation}

By combining Eq. \eqref{13} - Eq. \eqref{15} together, we have the upper bound of $ E_{1} $, that is
\begin{equation}\label{16}
E_{1} = {\rm O} \left(r{{(n)}^{ - 1}}\right). 
\end{equation}

Similarly, the lower bound of $ E_{1} $ is
\[\begin{array}{l}
E_{1} > \sum\limits_{x=1}^{\frac{1}{{r(n)}}} {x\sum\limits_{l = 1}^{4x} {\sum\limits_{{v_k} \in {s_l}} {\frac{{q_k^{ - \epsilon }{\sigma _{q - 1,N}}(Q) - q_k^{ - 2\epsilon }{\sigma _{q - 2,N}}(Q)}}{q\cdot{{\sigma _{q,N}}(Q)}}} } } \\
\quad= {\rm{upper\ bound - }}\sum\limits_{x=1}^{\frac{1}{{r(n)}}} {x\sum\limits_{l = 1}^{4x} {\sum\limits_{{v_k} \in {s_l}} {\frac{{q_k^{ - 2\epsilon }{\sigma _{q - 2,N}}(Q)}}{q\cdot{{\sigma _{q,N}}(Q)}}} } } .
\end{array}\]

It turns out that the second term in the lower bound is
\begin{equation}\label{17}
\sum\limits_{x=1}^{\frac{1}{{r(n)}}} {x\sum\limits_{l = 1}^{4x} {\sum\limits_{{v_k} \in {s_l}} {\frac{{q_k^{ - 2\epsilon }{\sigma _{q - 2,N}}(Q)}}{q\cdot{{\sigma _{q,N}}(Q)}}} } } \equiv \Theta \left(n^{-1}\cdot r{{(n)}^{ - 1}}\right).
\end{equation}

The order in Eq. \eqref{17} is negligible compared with the upper bound in Eq. \eqref{16}, so the order of $ E_{1} $ is solved. $\hfill\blacksquare$ 

$ \mathbf{Lemma\ 5.} $ When the degree of the source node is greater than $ q_{0} $, i.e. $ q>{q_0} $, the average number of hops $ E_{2} $ is
\begin{equation}\label{18}
E_{2} =\Theta \left(r{{(n)}^{ - 1}}\right). 
\end{equation}

Proof: Similar to the case $ E_{1} $, $ E_{2} $ is given as
\begin{equation}\label{19}
{E_2} = \sum\limits_{q = {q_0} + 1}^n {\frac{{{q^{ - \gamma }}}}{{\sum_{b = 1}^n {{b^{ - \gamma }}} }}}  \cdot \sum\limits_{x=1}^{\frac{1}{{r(n)}}} {x\sum\limits_{l = 1}^{4x} {\sum\limits_{{v_k} \in {s_l}} {\frac{{q_k^{ - \epsilon }\sigma _{q - 1,N - 1}^{\overline k }(Q)}}{q\cdot {{\sigma _{q,N}}(Q)}}} } } .
\end{equation}

Since $ N $ is large enough and the degrees of $ q $ direct social contacts are independent and identically distributed, the law of large numbers can work here. Let $ {{\rm{X}}_{{i_j}}} = q_{{i_j}}^{ - \epsilon },{{\rm{Y}}_{{i_j}}} = \log {X_{{i_j}}} $ and $ \overline Y  $ denotes the mean of $ Y_{i_{j}} $ , then we have
\begin{equation}\label{20}
\begin{array}{l}
\frac{{q_k^{ - \epsilon }\sigma _{q - 1,N - 1}^{\overline k }(Q)}}{q\cdot{{\sigma _{q,N}}(Q)}} \equiv \frac{{\sum_{1 \le {i_1} \le  \cdot  \cdot  \cdot  \le {i_q} \le N,\exists m,{i_m} = k} {\prod_{{\rm{j}} = {\rm{1}}}^{\rm{q}} {{{\rm{X}}_{{{\rm{i}}_{\rm{j}}}}}} } }}{q\cdot{\sum_{1 \le {i_1} \le  \cdot  \cdot  \cdot  \le {i_q} \le N} {\prod_{{\rm{j}} = {\rm{1}}}^{\rm{q}} {{{\rm{X}}_{{{\rm{i}}_{\rm{j}}}}}} } }}\\
\equiv\frac{{\sum_{1 \le {i_1} \le  \cdot  \cdot  \cdot  \le {i_q} \le N,\exists m,{i_m} = k} {\exp (\sum_{{\rm{j}} = {\rm{1}}}^{\rm{q}} {{{\rm{Y}}_{{{\rm{i}}_{\rm{j}}}}})} } }}{q\cdot{\sum_{1 \le {i_1} \le  \cdot  \cdot  \cdot  \le {i_q} \le N} {\exp (\sum_{{\rm{j}} = {\rm{1}}}^{\rm{q}} {{{\rm{Y}}_{{{\rm{i}}_{\rm{j}}}}})} } }}\\
\equiv \frac{{\sum_{1 \le {i_1} \le  \cdot  \cdot  \cdot  \le {i_q} \le N,\exists m,{i_m} = k} {{\rm{        }}\exp (q\overline {\rm{Y}} )} }}{q\cdot{\sum_{1 \le {i_1} \le  \cdot  \cdot  \cdot  \le {i_q} \le N} {{\rm{    }}\exp (q\overline {\rm{Y}} )} }}\\
\equiv \frac{{\left( {\begin{array}{*{20}{c}}
			{N - 1}\\
			{q - 1}
			\end{array}} \right)}}{q\cdot{\left( {\begin{array}{*{20}{c}}
			N\\
			q
			\end{array}} \right)}} = \frac{1}{N} = \Theta (n^{-1}).
\end{array}
\end{equation}

Besides, the probability that the degree of the source node is greater than $ q_{0} $ is 
\begin{equation}\label{21}
P(q > {q_0}) = \sum\limits_{q = {q_0} + 1}^n {\frac{{{q^{ - \gamma }}}}{{\sum_{b = 1}^n {{b^{ - \gamma }}} }}}  = \Theta (1).
\end{equation}

Then Eq. \eqref{19} can be simplified by Eq. \eqref{20} - Eq. \eqref{21}, namely:
\begin{equation}\label{22}
E_{2} \equiv \sum\limits_{x=1}^{\frac{1}{{r(n)}}} {x\sum\limits_{l = 1}^{4x} {\sum\limits_{{v_k} \in {s_l}} {\frac{1}{n}} } }\equiv\Theta \left(r{{(n)}^{ - 1}}\right).
\end{equation}

Therefore, the order of $ E_{2} $ is obtained. $\hfill\blacksquare$

Now we can prove Theorem 1, from Lemma 4 and Lemma 5, we have
\begin{equation}\label{23}
E[X] = {E_1} + {E_2} = \Theta \left(r{{(n)}^{ - 1}}\right).
\end{equation}

Together with Lemma 2, we can obtain the results in Theorem 1. $\hfill\blacksquare$ 

\subsection{The Case of Power-law Distributed Destinations}
In the second case, we assume that the probability that the source node communicates with one of its direct social contacts is proportional to $ d^{-\beta} $, where $ d $ refers to the distance between the source node and the destination node, and $ \beta $ indicates that the closer social contacts have more opportunities to communicate with the source node. Different from the case of uniformly distribued destinations, fractal wireless networks achieve another maximum throughput in this situation, which is clarified in Theorem 2 and proofed in the remainder of the subsection.

$ \mathbf{Theorem\ 2.} $ For a fractal wireless network with $ n $ nodes satisfying the conditions below: 1) the direct social contacts are selected according to the joint probability distribution $ P({k_1},{k_2}) = \frac{{k_1^{ - (\gamma  - 1)}k_2^{ - \epsilon }}}{{{M_{\gamma ,\epsilon }}}} $; 2) the degree of each node follows the power-law degree distribution $ P(k) = \frac{{{k^{ - \gamma }}}}{{\sum_{k = 1}^n {{k^{ - \gamma }}} }} $; 3) the destination node is chosen by the power-law distribution $ P({v_t} = {v_k}|{v_k} \in {\bf{C}}) = \frac{{{d^{ - \beta }}}}{{\sum_{j = 1}^q {d_j^{ - \beta }} }} $.  Then the maximum capacity $ \lambda_{max} $ of the fractal wireless network with direct social interactions is
\[{\lambda _{\max }} = \left\{ {\begin{array}{*{20}{c}}
	\begin{aligned}
	&{\Theta\left(\frac{1}{{\sqrt {n \cdot \log n} }}\right),{\rm{     \ \quad \qquad\qquad 0}} \le \beta (\gamma ,\epsilon ) \le {\rm{2}}};\\
	&{\Theta\left(\frac{1}{{\sqrt {{n^{3 - \beta }} \cdot \log {n^{\beta  - 1}}} }}\right),{\rm{    \ \qquad 2}} < \beta (\gamma ,\epsilon ) \le {\rm{3}}};\\
	&{\Theta\left (\frac{1}{{\log n}}\right),{\rm{        \qquad\qquad\qquad\qquad }}\beta (\gamma ,\epsilon ) > 3}.
	\end{aligned}
	\end{array}} \right.\]

Proof: The proof of Theorem 2 is pretty similar to Theorem 1, so we just give some key lemmas in the derivation procedure. 

$ \mathbf{Lemma\ 6.} $ Let the source node $ v_{i} $ locate in the central square whose degree is $ q $, where $ q=1,2,\ldots,n $. $ v_{k} $ denotes the destination node whose degree is $ q_{k} $ and its distance from the source node is $ d_{k} $. The variables $ {q_1},{q_2}, \cdot  \cdot  \cdot ,{q_N} $ in $ Q = (q_1^{ - \epsilon },q_2^{ - \epsilon }, \cdot  \cdot  \cdot ,q_N^{ - \epsilon}) $ denote the degrees of $ N $ potential direct social contacts whose degrees are smaller than $ q $. Let $ D = (d_1^{ - \beta },d_2^{ - \beta }, \cdot  \cdot  \cdot d_q^{ - \beta }) $, and $ {d_j}\ (1\le j\le q) $ in $ D $ denotes the distance between the $ j $th social contact and the source node. Then the average number of hops can be given as 
\begin{equation}\label{24}
E[X] = \sum\limits_{q = 1}^n {\frac{{{q^{ - \gamma }}}}{{\sum_{b = 1}^n {{b^{ - \gamma }}} }}}  \cdot \sum\limits_{x=1}^{\frac{1}{{r(n)}}} {x\sum\limits_{l = 1}^{4x} {\sum\limits_{{v_k} \in {s_l}} {\frac{{q_k^{ - \epsilon }\sigma _{q - 1,N - 1}^{\overline k }(Q)}}{{{\sigma _{q,N}}(Q)}}} } }  \cdot \frac{{d_k^{ - \beta }}}{{{\sigma _{1,q}}(D)}}.
\end{equation}

$ \mathbf{Lemma\ 7.} $ When the degree of the source node is not greater than $ q_{0} $, i.e. $ q \le {q_0} $, the average number of hops $ E_{1} $ is
\begin{equation}\label{25}
E_{1} = \left\{ {\begin{array}{*{20}{c}}
	\begin{aligned}
	&{\Theta \left(r{{(n)}^{ - 1}}\right),{\rm{      \ \ \qquad 0}} \le \beta  \le {\rm{2}}};\\
	&{\Theta\left (r{{(n)}^{\beta  - 3}}\right),{\rm{     \qquad 2}} < \beta  \le {\rm{3}}};\\
	&{\Theta (1),{\rm{         \ \quad\qquad\qquad}}\beta  > 3}.
	\end{aligned}
	\end{array}} \right.
\end{equation}

$ \mathbf{Lemma\ 8.} $ When the degree of the source node is greater than $ q_{0} $, i.e. $ q>{q_0} $, the average number of hops $ E_{2} $ is
\begin{equation}\label{26}
E_{2} = \left\{ {\begin{array}{*{20}{c}}
	\begin{aligned}
	&{\Theta\left (r{{(n)}^{ - 1}}\right),{\rm{      \ \qquad 0}} \le \beta  \le {\rm{2}}};\\
	&{\Theta\left (r{{(n)}^{\beta  - 3}}\right),{\rm{    \qquad 2}} < \beta  \le {\rm{3}}};\\
	&{\Theta (1),{\rm{        \ \ \ \qquad\qquad }}\beta  > 3}.
	\end{aligned}
	\end{array}} \right.
\end{equation}

Combined with Lemma 2, we can obtain the result in Theorem 2. $\hfill\blacksquare$ 

\section{Discussions}
After mathematically obtaining the maximum achievable capacity in Section III, we display the results in an intuitive way. Fig. 3 illustrates the simulation diagrmas of the maximum capacity in two different cases.
\begin{figure}[htbp]
		\centering
		\includegraphics[scale=0.4]{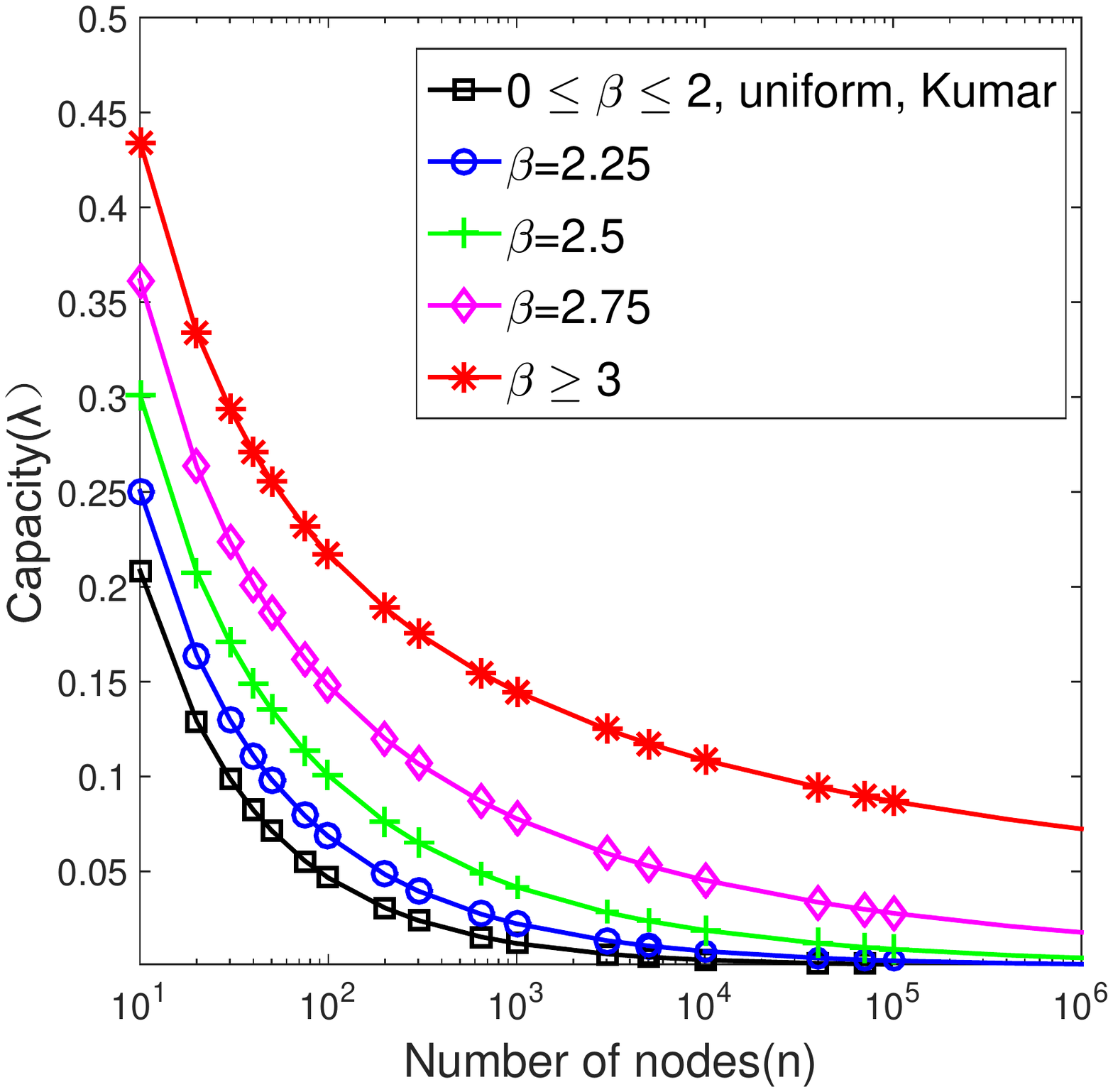}
		\caption{The maximum capacity in Theorem 1 with different values of $ \beta $.}
	\end{figure}

By Theorem 2, we find that the capacity varies for different values of $ \beta $. When $ \beta $ has a relatively small value, there is no obvious increase in the throughput compared with the classical result achieved by Kumar in \cite{Gupta2000The}. Particularly, when $ \beta=0 $, the situation is reduced to the one with uniformly distributed destinations as stated in Section 3.1. As $ \beta $ increases to $ 2\sim 3 $, the throughput leads to an exponential growth. After it exceeds $ 3 $, the throughput reaches up to $ \Theta\left (\frac{1}{{\log n}}\right) $, which exhibits remarkable improvement compared with $ \Theta\left (\frac{1}{{\sqrt {n\log n} }}\right) $ in \cite{Gupta2000The}. This increase can be explained by the reduction in the average number of hops for each transmission between the source node and the destination node, which is depicted in Fig. 4.
\begin{figure}[htbp]
	\centering
	\includegraphics[scale=0.4]{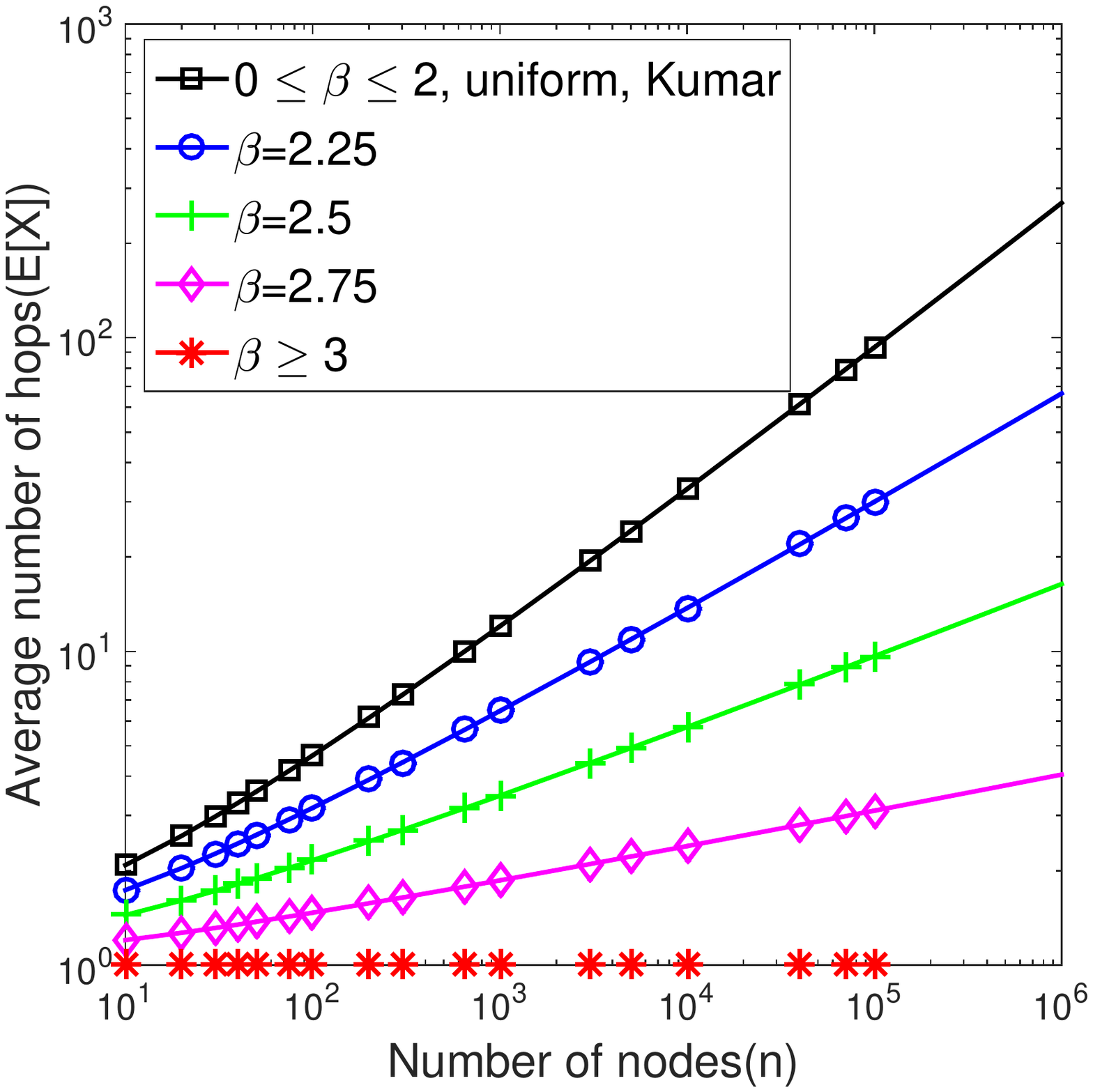}
	\caption{The average number of hops with different values of $ \beta $.}
\end{figure}

As shown in Fig. 4, when $ \beta $ varies between 0 and 2, the average number of hops does not decrease distinctly compared with the case where the destination nodes are chosen uniformly. When $ \beta $ exceeds 2, the source node prefers to communicate with the closer direct social contacts, which leads to the exponentially growth of the maxium throughput. When $ \beta $ is greater than 3, only $ \Theta (1) $ average hops are taken for each transmission, which raises the maximum capacity up to  $ \Theta\left (\frac{1}{{\log n}}\right) $ finally.  

\section{Conclusion and Future Works}
In this paper, we study the capacity of fractal wireless networks with direct social interactions. It has been proved that if the source node communicates with one of its direct social contacts randomly, the maximum capacity in Theorem 1 corresponds with the classical result $ \Theta\left(\frac{1}{\sqrt{n\log n}}\right) $ achieved by Kumar \cite{Gupta2000The}. On the other hand, if the two nodes with distance $ d $ communicate according to the probability $ d^{-\beta} $, the maximum capacity in Theorem 2 is
\[ {\lambda _{\max }}= \left\{ {\begin{array}{*{20}{c}}
			\begin{aligned}
			&{\Theta\left (\frac{1}{{\sqrt {n \cdot \log n} }}\right),{\rm{     \qquad\qquad\qquad 0}} \le \beta(\gamma,\epsilon)  \le {\rm{2}}};\\
			&{\Theta\left (\frac{1}{{\sqrt {{n^{3 - \beta }} \cdot \log {n^{\beta  - 1}}} }}\right),{\rm{    \qquad\quad 2}} < \beta(\gamma,\epsilon)  \le {\rm{3}}};\\
			&{\Theta\left (\frac{1}{{\log n}}\right),{\rm{        \ \qquad\qquad\qquad\qquad }}\beta(\gamma,\epsilon)  > 3}.
			\end{aligned}
			\end{array}} \right. \]

We notice that $ \beta $ has a straight-forward effect on the maximum capacity of fractal wireless networks with direct social interactions. Besides, as $ \beta $ might be influenced by the properties of fractal wireless networks, the maximun capacity is related to the degree distribution exponent $ \gamma $ and the correlation exponent $ \epsilon $. Although they are not included in the results explictly, it is reasonable that they have influence on the capacity of the fractal wireless networks. In other words, the degree distribution exponent $ \gamma $ and the correlation exponent $ \epsilon $ would determine the maximum throughput to some extent. 

What is the explict relationship between $ \beta $ and the parameters $ \gamma $ and $ \epsilon $? How do $ \beta $ and $ \epsilon $ shape the final capacity? We leave all these open issues in the future works.

\bibliographystyle{IEEEtran}
\bibliography{IEEEfull,reference}

\end{document}